%% file: sn-article.tex
\documentclass[pdflatex,sn-mathphys-num]{sn-jnl}

\usepackage{graphicx}%
\usepackage{multirow}%
\usepackage{amsmath,amssymb,amsfonts}%
\usepackage{amsthm}%
\usepackage{mathrsfs}%
\usepackage[title]{appendix}%
\usepackage{xcolor}%
\usepackage{textcomp}%
\usepackage{manyfoot}%
\usepackage{booktabs}%
\usepackage{algorithm}%
\usepackage{algorithmicx}%
\usepackage{algpseudocode}%
\usepackage{adjustbox}
\usepackage{listings}%
\usepackage{tikz}
\usepackage{bm}
\usetikzlibrary{positioning,arrows,shadows,calc,trees}

\theoremstyle{thmstyleone}%
\theoremstyle{thmstyletwo}%

\theoremstyle{thmstylethree}%

\begin{document}

\title[Predicting Thermodynamics of Liquid Water from Time Series Analysis]{Predicting the Thermodynamics of Liquid Water from Time Series Analysis}

\author[1,2]{\fnm{Ma{{\l}}gorzata J.} \sur{Zimo\'{n}}}\email{malgorzata.zimon@uk.ibm.com}

\author[1,3]{\fnm{Fausto} \sur{Martelli}}\email{fausto.martelli@cnr.it}

\affil[1]{\orgname{IBM Research Europe}, \orgaddress{\street{Keckwik Lane}, \city{Daresbury}, \postcode{WA4 4AD}, \country{United Kingdom}}}

\affil[2]{\orgdiv{Department of Mathematics}, \orgname{University of Manchester}, \orgaddress{\street{Alan Turing Building, Oxford Rd.}, \city{Manchester}, \postcode{M13 9PL}, \country{United Kingdom}}}
 \affil[3]{\orgdiv{Institute for Complex Systems}, \orgname{CNR}, \orgaddress{\street{Piazzale Aldo Moro 5}, \city{Rome}, \postcode{00185}, \country{Italy}}}


\abstract{Thermodynamics, introduced over two centuries ago, remains foundational to our understanding of physical, chemical, biological, and engineering systems. Its principles are traditionally grounded in the statistical mechanics framework, which explains macroscopic behavior from microscopic states. In this work, we propose an alternative approach that interprets thermodynamic behavior through the lenses of time series analysis, an approach commonly used in other fields, including finance, climate, and signal processing. We perform classical molecular dynamics simulations of liquid water, the most complex, anomalous, and important substance known, over a wide range of its phase diagram. By examining the temporal evolution of the hydrogen bond network (HBN) topology, we demonstrate that the dynamics of microscopic topological motifs populating the HBN encode the system’s macroscopic thermodynamic behavior. Furthermore, our approach enables the prediction of thermodynamic properties in regions beyond those directly sampled in our simulations. We achieve this result by leveraging artificial intelligence to uncover patterns in temporally resolved data that are often lost through conventional averaging. This work offers new insights into the fundamental behavior of water and network-forming materials more broadly, establishing a new paradigm for understanding material properties beyond the classical confines of statistical mechanics.}




\maketitle

\section{Introduction}\label{sec1}
Water possesses the most complex phase diagram among all known pure substances~\cite{salzmann2019advances}, comprising over twenty crystalline phases and several amorphous states~\cite{amann2016colloquium,rosu2023medium}. This exceptional poly(a)morphism stems from the unique geometry of the water molecule and the flexibility of its hydrogen bonds (HBs), which can readily reorganize in response to variations in pressure and temperature. HBs form an extended link of connectivity that plays a central role in governing water’s behavior across its diverse phases. 
In glassy water, molecules are kinetically arrested in disordered configurations akin to those of a liquid, but each water molecule is almost perfectly tetrahedrally coordinated. The absence of dangling bonds creates a hydrogen bond network (HBN) that efficiently absorbs large-scale density fluctuations, resulting in low compressibility~\cite{martelli2017large,martelli2022steady,formanek2023molecular}. Pervasive rearrangements in the HBN instead take place in correspondence with pressure-induced phase transitions between amorphous phases~\cite{mishima1985apparently,mishima1994reversible,giovambattista2016potential,martelli2018searching,mollica2022decompression} and are responsible for the associated sharp peaks in compressibility~\cite{formanek2023molecular,gutierrez2024link}. In plastic ice VII--a recently identified high-pressure phase in which water molecules occupy fixed lattice positions, as in conventional ice VII, but retain rotational freedom akin to that in a liquid~\cite{rescigno2025observation}-- the rearrangements in the HBN dictate the ductility~\cite{toffano2022temperature,martelli2023electrolyte}. \newline
However, it is in the liquid phase that the role of the HBN becomes particularly pronounced. Liquid water exhibits the highest number of thermodynamic and dynamic anomalies among pure substances~\cite{gallo2016water}. Liquid water displays a density profile characterized by a maximum and a minimum at lower temperatures~\cite{liu2007observation}. The maxima in water’s thermodynamic response functions --associated with the Widom lines~\cite{kim2017maxima}~\footnote{Historically, the Widom line is the line of maxima in heat capacity. However, for the sake of simplicity, we here consider the line of maxima of isothermal compressibility as a second Widom line.}-- are linked to spatial arrangements of water molecules~\cite{russo2014understanding} and emerge when the HBN adopts a configuration characterized by an approximately equal number of pentagonal and hexagonal hydrogen-bonded motifs, or rings, pointing towards a possible link between HBN topology and thermodynamic response functions~\cite{martelli2019unravelling}. Within this network, water molecules are represented as vertices and hydrogen bonds as edges, and an $n$-member ring represents a cyclic path consisting of a sequence of vertices and edges formed by $n$ hydrogen-bonded water molecules~\cite{martelli2019unravelling,formanek2020probing}. The structural arrangement arising in correspondence with the maxima of the thermodynamic response functions corresponds to a highly frustrated state with an equal amount of pentagonal and hexagonal rings populating the HBN. Hexagonal rings promote local crystalline-like order, while pentagonal motifs frustrate against it, thereby supporting liquid-like behavior. A similar interplay between order and frustration was later observed in supercooled liquid silicon~\cite{goswami2023analysis}, another prototypical anomalous liquid~\cite{sastry2003liquid,ganesh2009liquid,beye2010liquid,vasisht2011liquid}. Sciortino et al. later inspected the properties of the HBN of liquid water at conditions close to the liquid-liquid critical point~\cite{foffi2021structure,foffi2021structural}, also examining how rings interpenetrate and generate topological invariants~\cite{neophytou2022topological}. Building on these insights, Guti\'errez Fosado et al.~\cite{gutierrez2024link} proposed that densification in network-forming materials may be driven by the formation of interpenetrating topological motifs in the network~\cite{gutierrez2024link}. A similar conclusion was then drawn by Neophytou et al. upon investigating DNA-functionalized nanoparticles~\cite{neophytou2024hierarchy}. Finally, in a combined experimental and computational study, Palombo et al.~\cite{palombo2025topological} were able to connect the mechanical properties of DNA nanostar hydrogels with the topology of the network~\cite{palombo2025topological}. There is, therefore, a growing corpus of evidence that thermodynamics is intimately linked to the topology of the network of bonds in network-forming materials. 

\begin{figure}[h]
\centering
\includegraphics[width=0.9\textwidth]{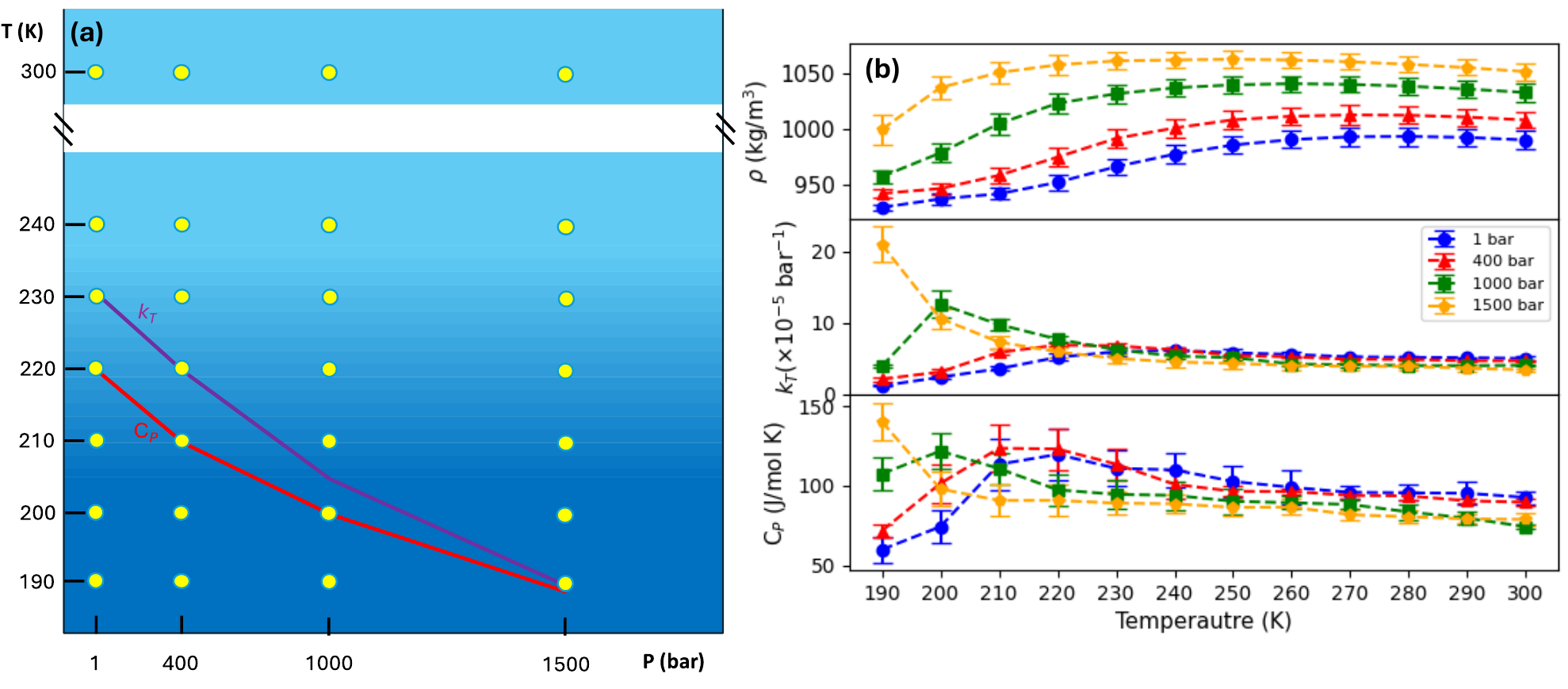}
\caption{Panel (a): sketch of the phase diagram of liquid water investigated in this work. Yellow circles represent the simulated state points. The red line represents the Widom line with the maxima in $C_P$. The purple line represents the line of maxima in $k_T$. Panel (b): profiles of the thermodynamic quantities as a function of the temperature at 1 bar (blue circles), 400 bar (red triangles), 1000 bar (green squares) and 1500 bar (yellow diamond).}\label{fig:task1}
\end{figure}

In this work, we focus on water and prove that such a link indeed exists. To achieve this goal, we move beyond the realm of statistical mechanics and enter the domain of time series analysis by considering the evolution in time of the geometrical motifs populating the HBN of liquid water over a wide range of thermodynamic conditions (sketched in Fig.~\ref{fig:task1}). In doing so, we gain access to the large amount of information that is lost in running averages, and that may hinder important particulars about the properties of the system. We consider three thermodynamic response functions, namely the density $\rho$, isothermal compressibility $k_T$, and isobaric heat capacity $C_P$. Although response functions are mathematically related --for example, compressibility and specific heat are linked through the coefficient of thermal expansion-- we argue that examining all of them, rather than focusing on just one, is essential to demonstrate the strength and validity of our results. Given the exploratory nature of this investigation, we aim to minimize potential errors arising from numerical differentiation and other unforeseen issues.

We perform classical molecular dynamics (MD) simulations of water described with the TIP4P/2005~\cite{abascal2005general} interaction potential and at equilibrium conditions (see the Methods Section for further computational details). At each thermodynamic point, we investigate the time evolution of the HBN; we compute the evolution in time of $n$-member rings with $n\in[3,12]$ and feed the signals to a model consisting of a bidirectional, multi-layer gated recurrent unit (GRU), a type of recurrent neural network (RNN) well-suited for sequential data. GRU utilises special gates, the update and reset gates, to reduce gradient dispersions. Compared to another commonly used variant of RNN, long short-term memory (LSTM)~\cite{hochreiter1997long}, the GRU mechanism introduced by~\citet{cho2014properties} has a simpler architecture, and is more efficient. Bidirectionality enhances the architecture by integrating input information from both past and future time steps. In our model, the final states of GRUs from the last layers are fed into a feedforward network to obtain the final prediction of a thermodynamic quantity. With this setup, we can learn and predict macroscopic behavior directly from microscopic topological descriptors. \newline
RNNs have made breakthroughs in natural language processing and speech recognition~\cite{cho2014learning,sutskever2014sequence,graves2013speech,graves2014towards}, and are relatively new in numerical simulations for materials. RNNs have been adopted to accelerate fine-elements and multiscale simulations for plasticity~\cite{ghavamian2019accelerating,wu2020recurrent,abueidda2021deep,borkowski2022recurrent}. Other studies have considered molecular dynamics (MD) simulations as time series that can be artificially augmented using RNNs~\cite{eslamibidgoli2019recurrent,tsai2020learning,wang2020accelerated,andrews2022forecasting,kadupitiya2022solving,fu2022simulate,winkler2022high,kadupitiya2022solving,lopez2023gpcr,dayhoff2025mlmd}. Other methods took a more ``direct'' approach and deterministically forecast changes in current positions and momenta at future times~\cite{zheng2021learning,klein2023timewarp,schreiner2023implicit,xu2024equivariant,jing2024generative,thiemann2025force,bigi2025flashmd}. These strategies, however, consider the whole MD trajectory as a time series. We go beyond existing approaches by demonstrating that microscopic physical quantities can be examined through the lenses of time series analysis. Our method represents a paradigm shift—from focusing on averaged quantities to analyzing detailed time series—and demonstrates that this approach enables access to a large body of information that is typically generated and discarded during numerical experiments.

\section{Results}\label{sec2}
We set up two tasks. The first task is a regression problem in which we use most of the signal to train the model, and the remaining part to predict the three thermodynamic quantities. The second task (also a regression task) is a prediction problem in which we train the model on three isobars and predict the thermodynamic quantities for the fully unseen fourth isobar. This task is therefore more complex, as we want to predict thermodynamic quantities on 12 unseen points in the phase diagram (from 300 K to 190 K at step of 10 K), and it is also more challenging because the model learns from only 36 signals. In the training, we use fixed-length time segments to optimize the model. As the signals are statistically nonstationary, this poses another difficulty. Ideally, different window lengths should be considered for each point depending on the system's dynamics. Moreover, we also aim at predicting response functions on pressures outside the bounds of the training set, making this task even more challenging. Details of the models for the two tasks are reported in the Methods Section.

\subsection*{Task 1: Predicting patterns}
\begin{figure}
\centering
\centering
 \includegraphics[width=.9\textwidth]{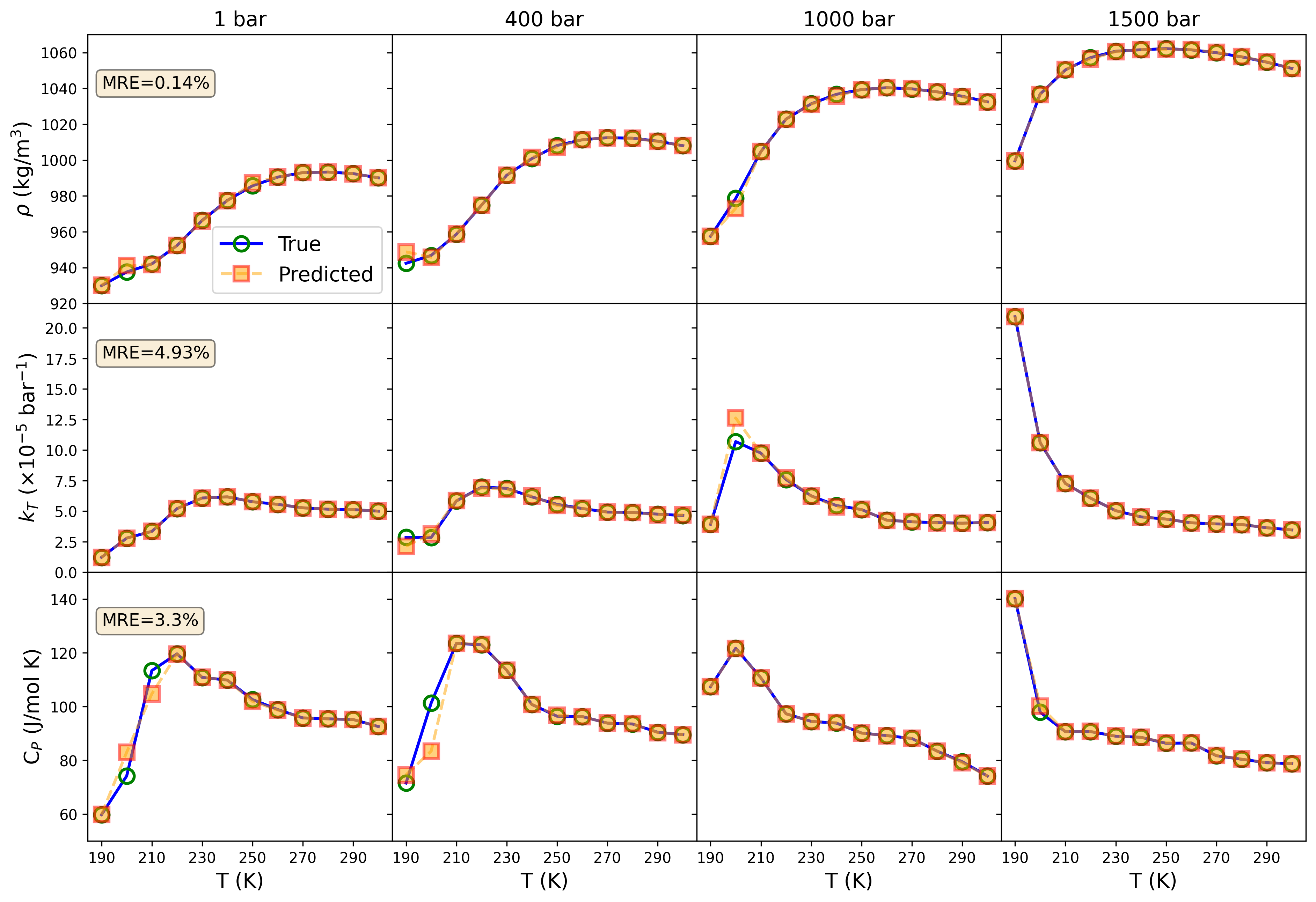}
\caption{Results for task 1. The values of each thermodynamic response function ($\rho$ upper row, $k_T$ mid row, $C_P$ lower row) are reported for each pressure. MD values are indicated as green, open circles, while predicted values are indicated as filled orange squares. The MREs reported in the plots at 1 bar are shared across isobars.}\label{fig:task_1_results}
\end{figure}

At each thermodynamic state point, the input signal is represented by a 10-dimensional vector of length $L=5000$, where $L$ denotes the number of snapshots collected and analyzed after equilibration. The dataset is divided into a training set (90\% of the signal) and a validation set (10\% of the signal), corresponding to 90 ns and 10 ns of simulation time, respectively. The model is trained to predict each thermodynamic quantity independently of pressure. \newline 
Figure~\ref{fig:schematic} shows a representative series of signals, ranging from the triangle to the dodecagon geometric motifs populating the HBN, that generate the response functions. Each state point is characterized by its unique 10-dimensional signal, and the model is trained by randomly selecting $N$ windows of $M=400$ snapshots each from the set of all signals. For validation, an observation window is slid across the final segment of the trajectory 100 consecutive times, covering the interval from 90 ns to 100 ns. In Fig.~\ref{fig:model} we report a sketch of the model. We train and validate three models, one for each thermodynamic response functions. Our approach is agnostic to the pressure and temperature because this information is never visible to the model. Figure~\ref{fig:task_1_results} displays the profiles of density ($\rho$, first row), isothermal compressibility ($k_T$, second row), and heat capacity at constant pressure ($C_P$, third row) at pressures of 1 bar, 400 bar, 1000 bar, and 1500 bar (from left to right). The predicted values (filled squares) closely follow the true values obtained from MD simulations (open circles), with near-perfect agreement across most conditions, as indicated by the mean relative error (MRE). Minor discrepancies appear at lower temperatures. This is expected given that the models are trained over the entire range of pressures. Moreover, the same training window lengths and trajectory durations are used at all conditions, despite the significantly slower dynamics observed under deep supercooling and elevated pressures. 
Overall, it is clear that when the model is given a random segment of the signal from an unseen time, it identifies the quantities that it represents. This experiment then shows that there is a unique pattern in the time series that relates the topology to thermodynamics.
\begin{figure}[h]
\centering
\includegraphics[width=0.9\textwidth]{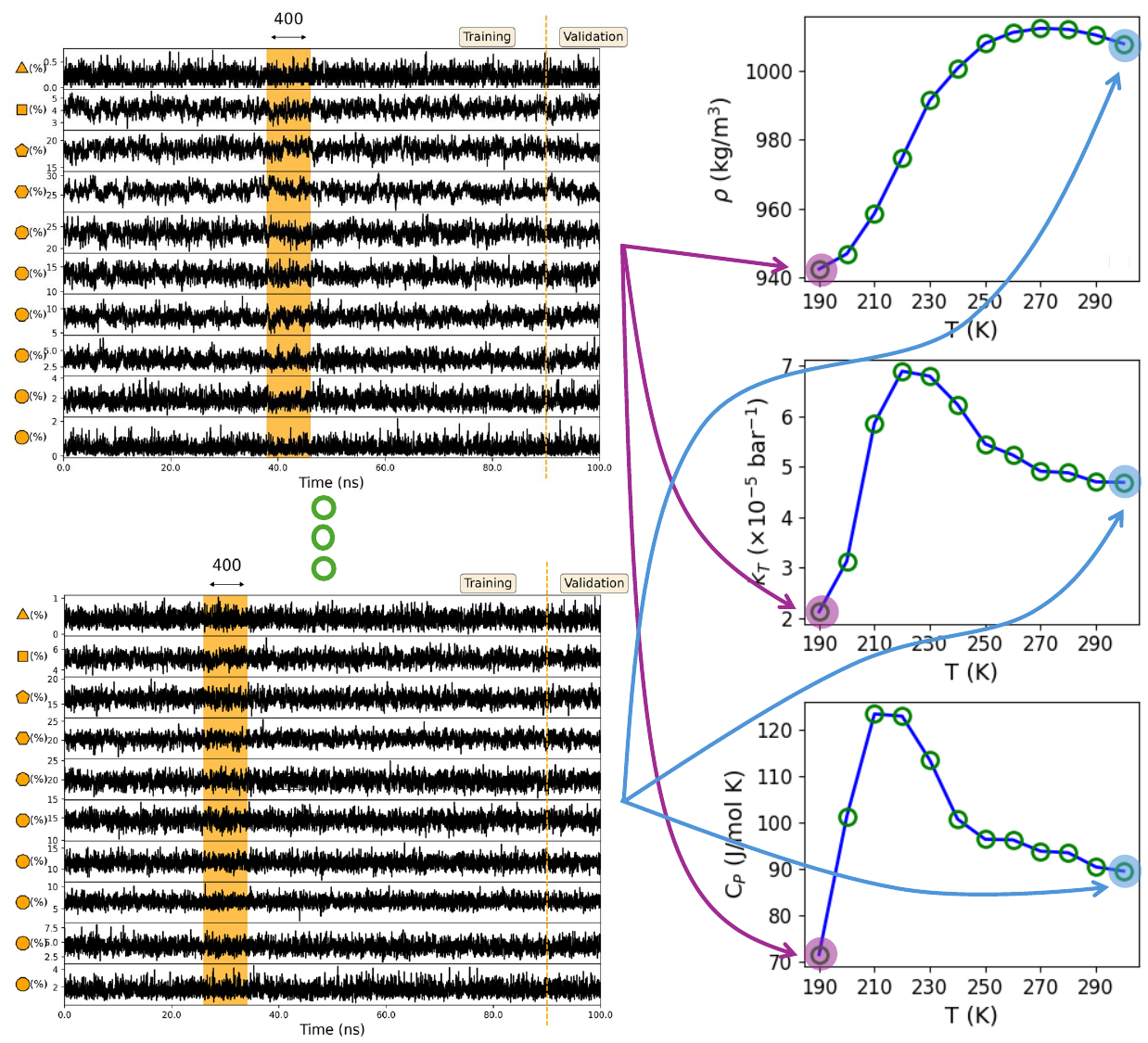}
\caption{Examples of signals (left) along an isobar generating response functions (right). Each signal is a 10-dimensional vector of length $L$=5000 snapshots corresponding to 100 ns. Each dimension corresponds to the evolution in time of the geometric motifs populating the HBN, shown on the left of the plots, from triangle to dodecagon (top to bottom). The signals are divided into a training set (left of the vertical dashed lines) and corresponding to  90\% of the signal, and a validation set (right of the vertical dashed line) corresponding to the remaining 10\% of the signal. The orange stripes represent sampling windows of length 400 snapshots.)}\label{fig:schematic}
\end{figure}
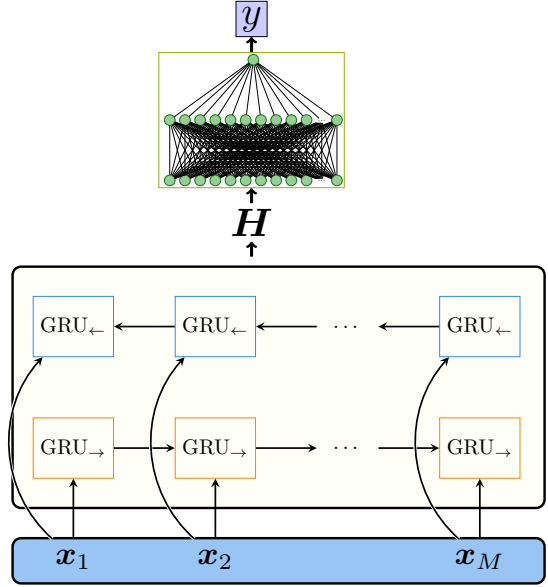
\begin{figure}
\centering
\begin{adjustbox}{scale=0.2}
    \input{tikz_images/model_img}
\end{adjustbox}  
\caption{Sketch of the model used to perform the study. The bidirectional multilayer GRUs are connected to a feedforward network with ReLU activation function and layer normalization. The output values are passed through a sigmoid function.}\label{fig:model}
\end{figure}

\subsection*{Task 2: Predicting unseen conditions}
\begin{figure}
\centering
\centering
 \includegraphics[width=.9\textwidth]{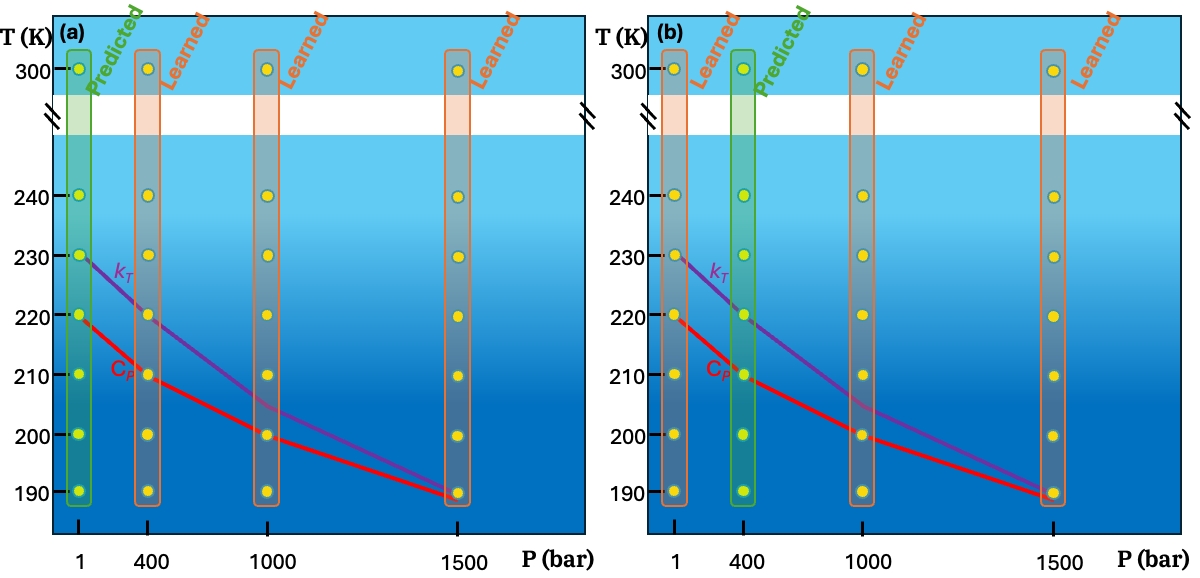}
 \includegraphics[width=.9\textwidth]{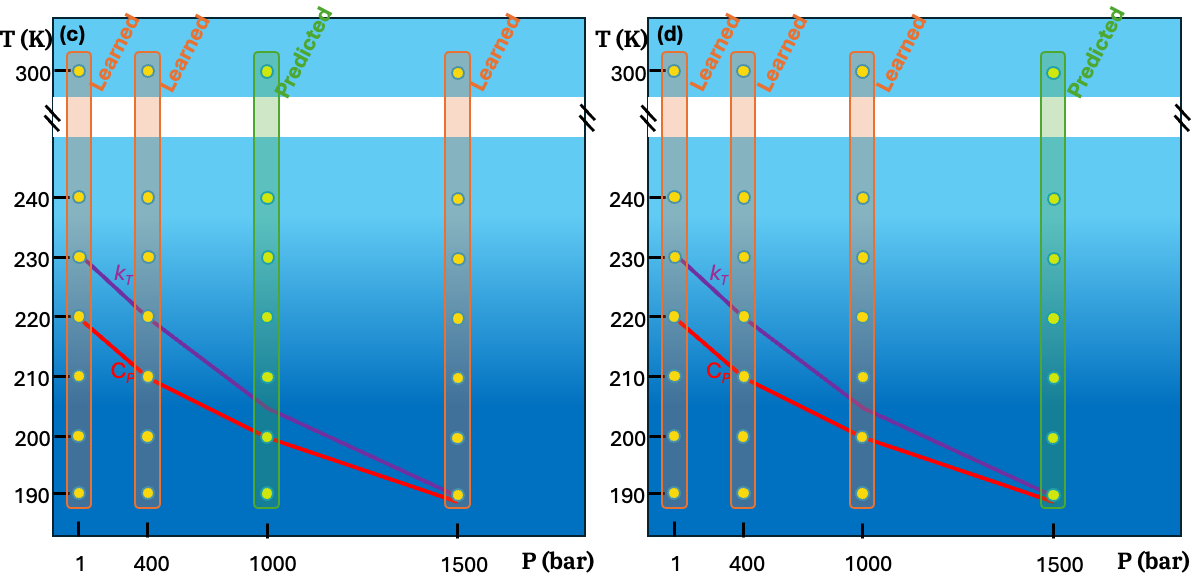}
\caption{Sketch of the phase diagram explored in task 2. In each panel, orange areas emphasize data points on isobars used to train the model, while green areas emphasize data points on which thermodynamic quantities are predicted. Panels (a) and (d) represent cases where predicted data fell outside the pressure values on which the model is trained.}\label{fig:task_2_PD}
\end{figure}
In the predictive task, we aim to train the models on three pressures and predict the thermodynamic quantities from the fourth. The study poses an additional challenge as the output data may lie outside the range of values present during the training. This is, for example, the case of the prediction of thermodynamic quantities on the isobars at 1 and at 1500 bars. In the first case, we train the model with data only from 400, 1000, and 1500 bar, and we aim at predicting the thermodynamic quantities on the isobar at 1 bar. In the second case, we train the model with data only from 1, 400, and 1000 bar, and we aim at predicting the thermodynamic quantities on the isobar at 1500 bar. A sketch of the possible scenarios is shown in Fig.~\ref{fig:task_2_PD}. As in Task 1, we train three models, one for each response function, including data from all considered pressures. Given the complexity of the task, we increase the training windows to 600 snapshots.

\begin{figure}
\centering
\centering
 \includegraphics[width=.9\textwidth]{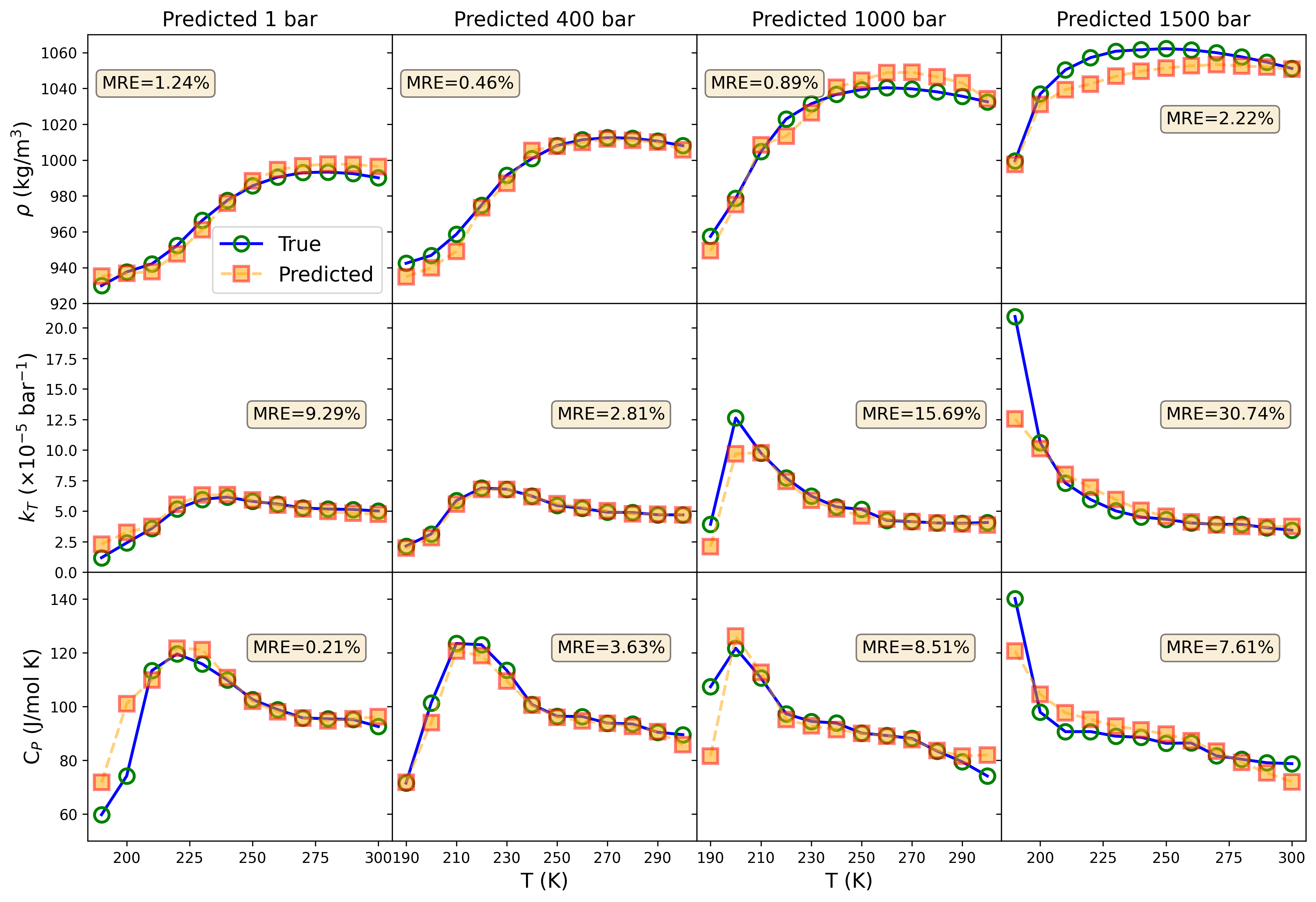}
\caption{Results for task 2. The values of each thermodynamic response function ($\rho$ upper row, $k_T$ mid row, $C_P$ lower row) are reported for each pressure. MD values are indicated as green, open circles, while predicted values are indicated as filled orange squares.}\label{fig:task_2_results}
\end{figure}

Figure~\ref{fig:task_2_results} presents the predicted thermodynamic conditions (solid squares) compared to the true values (open circles). 
Overall, the model accurately captures the behavior of all thermodynamic response functions, effectively reproducing both the non-monotonic features associated with the crossing of the Widom lines, as well as the smooth, monotonic profiles of $\rho$ and of $k_T$ and $C_P$ at 1500 bar. The first column of Fig~\ref{fig:task_2_results} reports the data predicted by the model at 1 bar with data learned at 400, 1000, and 1500 bar. The model correctly predicts all quantities over the entire isobar, as testified by the low MREs of 1.24\% for $\rho$, 9.29\% for the $k_T$, and 0.21\% for $C_P$. The model, therefore, successfully extrapolates thermodynamic conditions from higher to lower pressures. The second column of Fig~\ref{fig:task_2_results} reports the data predicted on the isobar at 400 bar with data learned at 1, 1000, and 1500 bar. As for the previous case, the model performs very well, as evidenced by the low MREs of 0.46\% for $\rho$, 2.81\% for $k_T$, and 3.63\% for $C_P$. The third column of Fig~\ref{fig:task_2_results} reports data predicted on the isobar at 1000 bar with data learned at 1, 400, and 1500 bar. The model, again, performs well, although not as well as in the previous cases. The MREs of 15.59\% for $k_T$ and of 8.51\% for $C_P$ are higher than in the previous cases. However, a close visual inspection shows that the main source of the errors comes from the two points at the lowest temperatures of 190 and 200 K. In these cases, the predicted points are slightly off, but the overall non-monotonous profile is still well captured by the model. The fourth column of Fig~\ref{fig:task_2_results} reports the data predicted on the isobar at 1500 bar with data learned from 1, 400, and 1000 bar. As for the previous case, the MREs (2.22\% for $\rho$, 30.74\% for $k_T$, and 7.61\% for $C_P$) are larger than at lower pressures, and the main contributors to the errors are the point at the lowest temperature of 190 K. Nonetheless, the profile of each thermodynamic quantity is qualitatively well capture. \newline 
The larger values of MREs for the predicted data at 1000 and 1500 bar should not be ascribed to a deviation in the validity of our approach. Rather, several factors can play a role in reducing the quality of the predictions on single points at low temperature and high pressure. Under these conditions, dynamics is sluggish, and a combination of longer trajectories, condition-adapted sampling windows, and optimized model parameters should fix the issue. Nevertheless, the experiment shows that the given dataset’s features can extrapolate/inform about unexplored regions of the thermodynamic map. The results again confirm the hypothesis that the topology is linked to the thermodynamics.

\section{Conclusions}
In this work, we demonstrate that the thermodynamic properties of liquid water are governed by the topology of the hydrogen-bond network (HBN). Departing from traditional statistical mechanics, we adopt a sequential data perspective to analyse this relationship. \newline
Our approach, based on gated recurrent units, successfully completes two tasks designed to prove the link between microscopic HBN topology and macroscopic quantities. The first is achieved with high fidelity. The model quantitatively recovers the density profiles at all isobars, and the monotonous profiles of $k_T$ and $C_P$ at 1000 bar, as well as the non-monotonous profiles of $k_T$ and $C_P$ upon crossing the Widom lines at 1, 400, and 1000 bar. The second, more challenging study also succeeds, though with somewhat less precision.  Qualitatively, the model predicts the profile of all response functions, underestimating $k_T$ and $C_P$ at 1000 and 1500 bar and 190 K, and overestimating $C_P$ at 1 bar and 200 K. However, considering (i) the limited data available, small by standard time-series analysis requirements, (ii) the consistent model parameters applied across all thermodynamic conditions, and (iii) the presence of an isobar exhibiting monotonic profiles of $k_T$ and $C_P$, these results represent a notable achievement. Further optimization of the model and expanded datasets (either running more and longer simulations, or using data augmentation from generative techniques such as, e.g., GANs/TimeGANs~\cite{goodfellow2014generative,yoon2019time} or variational autoencoders~\cite{kingma2013auto}) could improve performance. However, such refinements are outside the scope of this study. Finally, we emphasize that our AI-based approach is not the only possible. For instance, other techniques could be used and developed, including wavelets analysis (particularly powerful for studying non-stationary signals), or Fourier-based methods for signal processing. \newline 
As previously noted, the dataset used in this study is relatively small for data-intensive techniques like the one employed here. Therefore, the models are sensitive to the initialisation of the parameters and prone to overfitting. More complex architectures are not suitable in this setting. However, if more training samples were available, it would be beneficial to employ an attention layer~\cite{vaswani2017attention}. Such a mechanism enables a model to capture long-term dependencies and assess the relative importance of each component in a sequence compared to the other elements in a time series, possibly further improving the results.

Our study pioneers the treatment of physical observables as sequential data rather than as ensemble-averaged quantities. By harnessing artificial intelligence’s strength in extracting patterns from seemingly noisy data, often lost in averaging, this approach offers a new paradigm for understanding physics, chemistry, and materials science. 

\section{Methods}
\subsection{Molecular Dynamics Simulations}
Our study is based on classical molecular dynamics (MD) simulations of a system composed of N=1000 rigid water molecules described by the TIP4P/2005 interaction potential~\cite{abascal2005general} in the isobaric (NPT) ensemble. We perform extensive equilibrium simulations using the GROMACS 2018.1 software package~\cite{abraham2015gromacs}. Coulombic
and Lennard-Jones interactions are calculated with a cutoff distance of 1.1~nm, and long-range electrostatic interactions are treated using the Particle-Mesh Ewald (PME) algorithm. Temperatures and pressures are controlled using a Nos\'e-Hoover thermostat~\cite{nose2002molecular,hoover1985canonical} and a Parrinello-Rahman~\cite{parrinello1980crystal}. For the Nos\'e-Hoover thermostat, the period of the kinetic energy oscillations between the system and the reservoir is set to 1 ps; the time constant for the Berendsen barostat is set to 1 ps. Numerical integrations were performed with the Verlet algorithm~\cite{verlet1967computer} with a timestep of 2~fs. 

\subsection{Ring statistics}
Ring statistics is a theoretical tool that has proven to be instrumental in investigating the network topology in numerically simulated network-forming materials. The ring statistics is only one of many graph-based techniques to investigate network topologies and, in the case of water, it has helped in understanding the connections between water anomalies and thermodynamic response functions~\cite{martelli2019unravelling,formanek2020probing,foffi2021structural} as well as the properties of glassy water~\cite{martelli2022steady,formanek2023molecular}. We construct rings by starting from a tagged water molecule and recursively traversing the HBN until the starting point is reached, or the path exceeds the maximal ring size considered (12 water molecules in our case). The definition of hydrogen bond follows Ref.~\cite{luzar1996hydrogen}. According to this definition, two water molecules $A$ and $B$ are hydrogen bonded when the distance $d_{O_A-O_B}<3.5${\AA} and the angle $\widehat{O_BO_AH_A}<30^{\circ}$. We do not distinguish between the donor-acceptor character of the starting water molecule~\cite{martelli2020network}. 

\subsection{Machine learning model}
In each task, we use $N$ segments of the topological evolution $\bm{X} = \{\bm{x}_1,\cdots,\bm{x}_{M} \}$ as input data, in which $\bm{x}_i \in\mathbb{R}^{10}$. Inputs are normalised by dividing them by the maximum value. The labels are transformed using min-max normalisation. In the first task, depicted graphically in Fig.~\ref{fig:schematic}, we use the initial 90\% of the timesteps for training and validate if the model can predict the thermodynamic conditions from the remaining 10\% of the signals. During training, we randomly sample windows of length $M=400$ snapshots. Then, the data segments are passed through a model with four-layer bidirectional GRUs. Each GRU has a hidden dimension of $d_h = 128$. For every sample in the batch, the model outputs $\bm{H} \in \mathbb{R}^{2 d_h}$, which is a combination of the final states of the bidirectional GRUs from the last layer. Finally, a fully connected neural network ($f_{nn}$) is used to compute the output, $y$. We use a 2-layer network
\begin{equation}
    y = \sigma \left( f_{nn}(\bm{H}) \right) = \sigma \left( f(\bm{f}_1(\bm{H})) \right).
\end{equation}
In the above, the vector function is of the following form:
\begin{equation}
    \bm{f}_l(\bm{z}) \stackrel{\mathrm{def}}{=}\bm{g}_l \left( \bm{W}_l\bm{z} + \bm{b}_l \right),
\end{equation}
where $l$ defines a layer. In our study, the activation function $\bm{g}_l$ is ReLU~\cite{agarap2018deep}. The parameters $\bm{W}_l$ and $\bm{b}_l$ for each layer are learnt using Adam optimizer~\cite{kingma2014adam}. After the first layer, we apply layer normalization~\cite{ba2016layer} to stabilise the model. The final output is passed through a sigmoid function, $\sigma$, to avoid nonphysical estimates. The training takes 40000 epochs with a batch size of $N=128$; the hidden final state is then of size $\bm{H} \in \mathbb{R}^{N 2 d_h}$. The model, graphically shown in Fig.~\ref{fig:model}, is implemented using PyTorch~\cite{paszke2019pytorch}.

In the second task, we aim to extrapolate the patterns captured in the portion of the signals and use the trained model to predict thermodynamic conditions from unseen data. During the training, we sample $N=128$ segments of length $M=600$ from the signals obtained at 3 pressures and 12 temperatures at each pressure (36 10-dimensional signals), aiming to predict labels at the 4th pressure. In this case, two-layer bidirectional GRUs with the hidden dimension of $d_h = 64$ are used. The training is performed with $N=128$ samples in a batch and for 10000 epochs. A more complex model and training may be prone to overfitting.

We report the results with a percent error computed as
\begin{equation}
    \delta = \| \bm{y}_{\mathrm{pred}} - \bm{y}_{\mathrm{true}} \|_2 \times 100\% \mathbin{/} \| \bm{y}_{\mathrm{true}} \|_2,\label{eq:percent_error}
\end{equation}
where $ \bm{y}_{\mathrm{pred}}$ is a vector of predicted labels, $\bm{y}_{\mathrm{true}}$ denotes target values and $\| .\|_2$ is a vector norm. In the first task, we validate the model's ability to predict thermodynamic conditions using time windows of length 400 from the unseen last 500 snapshots. We slide the window 100 times and estimate the relative error for each segment. We then average the results over $N_v = 100$. Because of limited training data, the model is sensitive to the random seed. With few training samples, it is prone to getting trapped in local minima. Therefore, we test models obtained with $n_s = 10$ different seeds. The final result is the mean of all these cases:
\begin{equation}
    \frac{1}{n_s}\frac{1}{N_v}\sum_{j=1}^{n_s}\sum_{i=1}^{N_v}\delta_{ij}. \label{eq:mean_error}
\end{equation}
A similar approach is used in the second case. The model is validated using the time segments from unseen pressure. The validation window of size 600 snapshots is not moved along the entire signal, but over the length of 1500 snapshots characterized by the lowest error, generating $N_v=900$ samples. Manually selecting the window instead of randomly picking portions of the signal is justified, considering that the dynamics of the system are different at each thermodynamic condition. 
Again, models generated with $n_s = 10$ different seeds are used.



\backmatter


\bmhead{Acknowledgments}
We are grateful to Kyongmin Yeo for insightful discussions.

\bmhead{Authors contributions}
F.M. designed the research, F.M. performed molecular dynamics simulations and rings analysis, M.J.Z. performed neural networks analysis, F.M. and M.J.Z. discussed the results and wrote the paper.

\bigskip
\bibliography{sn-bibliography}

\end{document}

%% file: tikz_images/model_img.tex
\definecolor{mygreen}{HTML}{8DB600}
\definecolor{aqua}{HTML}{318CE7}

\begin{tikzpicture}
    \def \customY {-1};
    \node[draw=mygreen, ultra thick, anchor=south] (outputmlp) at (25, 8) {\input{tikz_images/mlp_img}};

    \node[scale=3, font=\Huge, below=of outputmlp.south, anchor=north] (grus_output) {$\bm{H}$};
     \node[draw=none, scale=4, below=of grus_output.south, ultra thick, anchor=north] (grus){\input{tikz_images/biGRU}};

    \node[draw=black, font=\Huge, fill=blue!20!white, anchor=south, above=of outputmlp.north, scale=3] (output) {$y$};
    
    \draw[->, black, line width=2mm] (outputmlp) -- (output);
    \draw[->, black, line width=2mm] (grus_output) -- (outputmlp.south);
    \draw[->, black, line width=2mm] (grus) -- (grus_output);
\end{tikzpicture}